\documentclass[10pt]{iopart}
\usepackage{graphicx}
\usepackage{amssymb}
\usepackage{iopams}
\usepackage{url}
\usepackage{lineno}

\newcommand {\figures}        {figures}
\newcommand {\ndof}           {\mbox{$N_{DOF}$ }}

\begin{document}
\title{Measurement of the energy and time resolution of a undoped CsI
  + MPPC array for the Mu2e experiment}
%
% author names and IEEE memberships
% note positions of commas and nonbreaking spaces ( ~ ) LaTeX will not break
% a structure at a ~ so this keeps an author's name from being broken across
% two lines.
% use \thanks{} to gain access to the first footnote area
% a separate \thanks must be used for each paragraph as LaTeX2e's \thanks
% was not built to handle multiple paragraphs
%

\author{O.~Atanova$^1$, M.~Cordelli$^2$, 
G.~Corradi$^2$, 
F.~Colao$^2$, 
Yu.I.~Davydov$^1$, 
R.~Donghia$^2$, 
S.~Di~Falco$^3$, 
S.~Giovannella$^2$, 
F.~Happacher$^2$, 
M.~Martini$^2$$^4$, 
S.~Miscetti$^2$, 
L.~Morescalchi$^3$$^5$, 
P.~Murat$^6$, 
G.~Pezzullo$^3$, 
A.~Saputi$^2$,
I.~Sarra$^2$, 
S.~R.~Soleti$^7$,
D.~Tagnani$^8$,
V.~Tereshchenko$^1$, 
Z.~Usubov$^1$}

%% \author{G.~Pezzullo$^{1}$, P.~Murat$^{2}$} 

%% \address{$^1$INFN of Pisa, Pisa, 56123 Italy}
%% \address{$^2$Fermi National Accelerator Laboratory, Batavia, IL 60510 USA}%

\address{$^1$Joint Institute for Nuclear Research, Dubna, Russia}
\address{$^2$Laboratori Nazionali dell'INFN, Frascati, Italy}
\address{$^3$INFN sezione di Pisa, Pisa, Italy}
\address{$^4$Universit\`a Guglielmo Marconi, Roma, Italy}
\address{$^5$Universit\`a di Siena, Siena, Italy}
\address{$^6$Fermi National Accelerator Laboratory, Batavia, Illinois, USA}
\address{$^7$University of Oxford, Oxford, United Kingdom}
\address{$^8$INFN sezione di Roma3, Roma, Italy}
\ead{ivano.sarra@lnf.infn.it, pezzullo@pi.infn.it}

\begin{abstract}
This paper describes the measurements of energy and time response and
resolution of a $3\times3$ array made of undoped CsI crystals coupled
to large area Hamamatsu Multi Pixel Photon Counters. The measurements
have been performed using the electron beam of the Beam Test Facility
in Frascati (Rome, Italy) in the energy range 80-120 MeV. The measured
energy resolution, estimated with the FWHM, at 100 MeV is
$16.4\%$. This resolution is dominated by the energy leakage due to the small
dimensions of the prototype. The time is reconstructed by fitting the
leading edge of the digitized signals and applying a digital constant
fraction discrimination technique. A time resolution of about 110 ps
at 100 MeV is achieved.
\end{abstract}

% Uncomment for PACS numbers
%
% Uncomment for keywords
\vspace{2pc}
\noindent{\it Keywords}:  Calorimetry, Timing, MPPC, undoped CsI crystal, Mu2e experiment
%
% Uncomment for Submitted to journal title message
%\submitto{\JPA}
%
% Uncomment if a separate title page is required
%\maketitle
% 
% For two-column output uncomment the next line and choose [10pt] rather than [12pt] in the \documentclass declaration
\ioptwocol
\pdfoutput=1

\section{Introduction}
The Mu2e experiment at FNAL~\cite{MU2ETDR} aims to observe the
charged-lepton flavor violating neutrinoless conversion of a negative
muon into an electron. The conversion results in a monochromatic
electron with an energy slightly below the muon rest mass (104.97
MeV). Two major elements of the detector system are the straw tube tracker and the
crystal calorimeter. The calorimeter~\cite{BIODOLA} should confirm
that the candidates reconstructed by the extremely precise tracker
system are indeed conversion electrons while performing a powerful
$\mu$/e particle identification. Moreover, it can provide a seed for the
track search and a high level trigger for the
experiment independently from the tracker system. The calorimeter
should also keep functionality in the highest irradiated areas where the background
delivers a dose of about 100 krad and a fluence of about 10$^{12}$
neutrons/cm$^2$ 1-MeV$_{\rm eq}$ in the hottest area, while working in
the presence of a 1 T axial magnetic field. These requirements translate
to a design of a calorimeter with large acceptance, good energy
resolution O(5$\%$) and a reasonable position (time) resolution better
than 1 cm (0.5 ns).

The baseline version of the calorimeter is composed of two disks of
inner (outer) radius of 37.4 (66.0) cm assembled from about 1350
undoped CsI crystals of $3.4\times3.4\times20$ cm$^3$ dimensions. Each
crystal is read out by two large area Silicon Photomultiplier (SiPM)
arrays. The undoped CsI emission spectrum peaks at 310 nm with a decay
time constant of about 30 ns~\cite{pdg2014}. The characteristics of
the undoped CsI are a reasonable match to the Mu2e requirements: light
yield (2000 photons/MeV), radiation length (1.9 cm) and Moli\`ere
radius (3.6 cm). Time resolution better than 0.5 ns has been recently
measured with an undoped CsI calorimeter at 100 MeV using a PMT-based
readout combined with a signal waveform digitization at 125
Msps~\cite{csi_calorimeter_kek_ps_nim_2005,koto_csi_nim_2015}.  The
tests discussed in this paper are motivated by the development of a
new generation of UV-extended SiPM~\cite{MPPCHAMAMATSU} from Hamamatsu
that improves the photo-detection efficiency (pde) close to the peak
in the wavelength emission spectrum of undoped CsI. This is obtained
by replacing the epoxy resin with a silicon protection layer. A
dedicated beam test was carried out during April 2015 at the Beam Test
Facility (BTF) in Frascati (Italy) where time and energy measurements
have been performed using a low energy electron beam, in the energy
range [80,120] MeV.

%%%%%%%%%%%%%%%%%%%%%%%%%%%%%%%%%%%%%%%%%%%%%%%%%%%%%%%%%%%%%%%%%%%%%%%%%%%%%%%%
\section{Experimental setup}
The calorimeter prototype consisted of nine $3\times3\times20$~cm$^3$
undoped CsI crystals wrapped in 150 $\mu$m of
Tyvek\textsuperscript{\textregistered}, arranged into a $3\times3$
matrix. Out of the nine crystals, two were produced by Filar
OptoMaterials~\cite{OPTOMATERIAL}, while the remaining 7 came from
ISMA~\cite{ISMA}. Each crystal was previously tested with a $^{22}$Na
source to determine its light output (LO) and longitudinal
response uniformity (LRU), with the results~\cite{RADHARD}:
\begin{itemize}
\item 
  a LO of about 90 pe/MeV, when read out with a UV-extended PMT R2059 by
  Hamamatsu~\cite{PMTR2059} coupled through an air gap;
\item
  a LRU corresponding to a LO variation at both ends of the crystals
  less than $\pm 6\%$.
\end{itemize}
Each crystal was coupled to a large area $12\times12$ mm$^2$ SPL TSV
SiPM (MPPC) from Hamamatsu~\cite{MPPCHAMAMATSU} by means of the
Rhodorsil 7 silicon paste~\cite{PASTE7}. SPL stands for ``silicon
protection layer'', while TSV for ``through silicon via'' and
indicates a new technique used for building the SiPM that is
characterized by a lower noise and a higher fill-factor. The operating
voltage was set at 55 V for each MPPC, about 3 V above the breakdown
voltage, corresponding to an average gain of $1.3\times10^6$ and a pde
of about $25-30\%$ at 300 nm. Each MPPC is composed of an array of 16
single SiPM, each one read out with its own anode. A front-end
electronics (FEE) board was developed to form an analog sum of the
pulses from several anodes. This board provides also a local HV
regulation and an amplification by a factor of 8. Photosensor signals
coming from the crystals and from a pair of scintillating counters
used for triggering were read out with 12 bit, 250 Msps waveform
digitizer boards, V1720 from CAEN~\cite{CAENDIGITIZER}.

\begin{figure}[h!]
  \centering
  \includegraphics[width=0.49\textwidth] {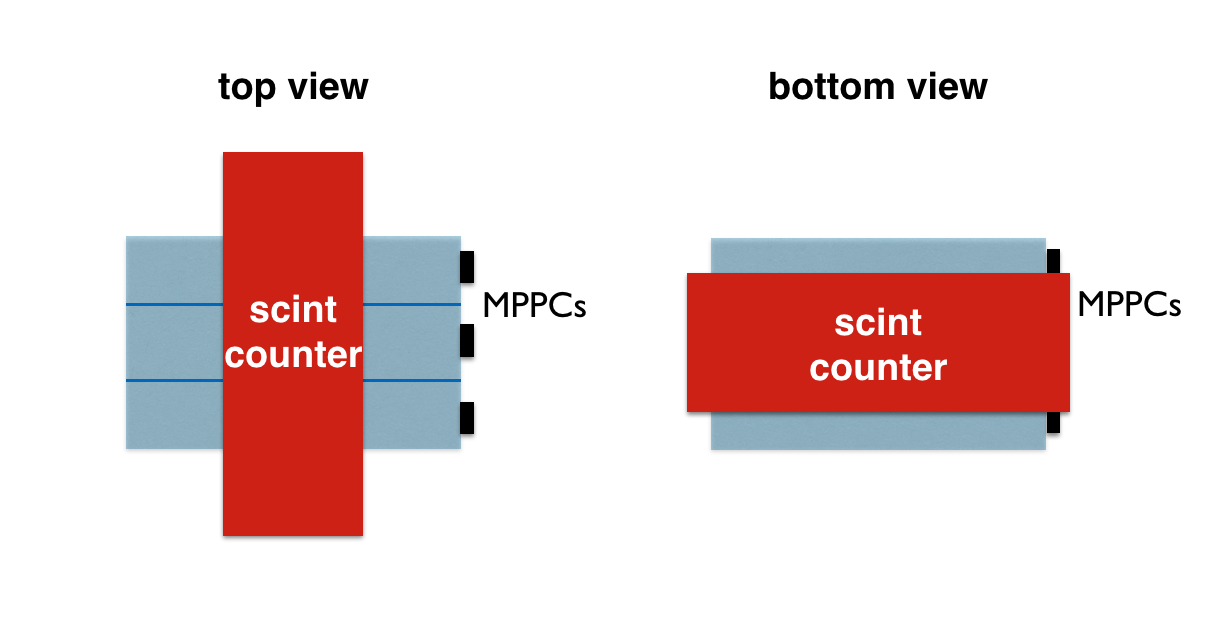}
  \caption{Cosmic ray counter orientation.}
  \label{fig:setup}
\end{figure}

The coincidence of the signals from two $5(\rm L)\times1(W)\times2(T)$~cm$^3$
plastic scintillating counters, crossed at 90 degrees, was used for
triggering on the incoming beam.
In addition, another
coincidence of the signals from two $10(\rm L)\times30(W)\times4(T)$ cm$^3$
scintillating counters, one above and one below the array as 
shown in Figure~\ref{fig:setup},  was used to provide a cosmic ray
trigger. %%Figure~\ref{fig:setup} shows the orientation of the cosmic
%% ray counters with respect to the prototype.

\begin{figure}[h!]
  \centering
  \includegraphics[width=0.49\textwidth] {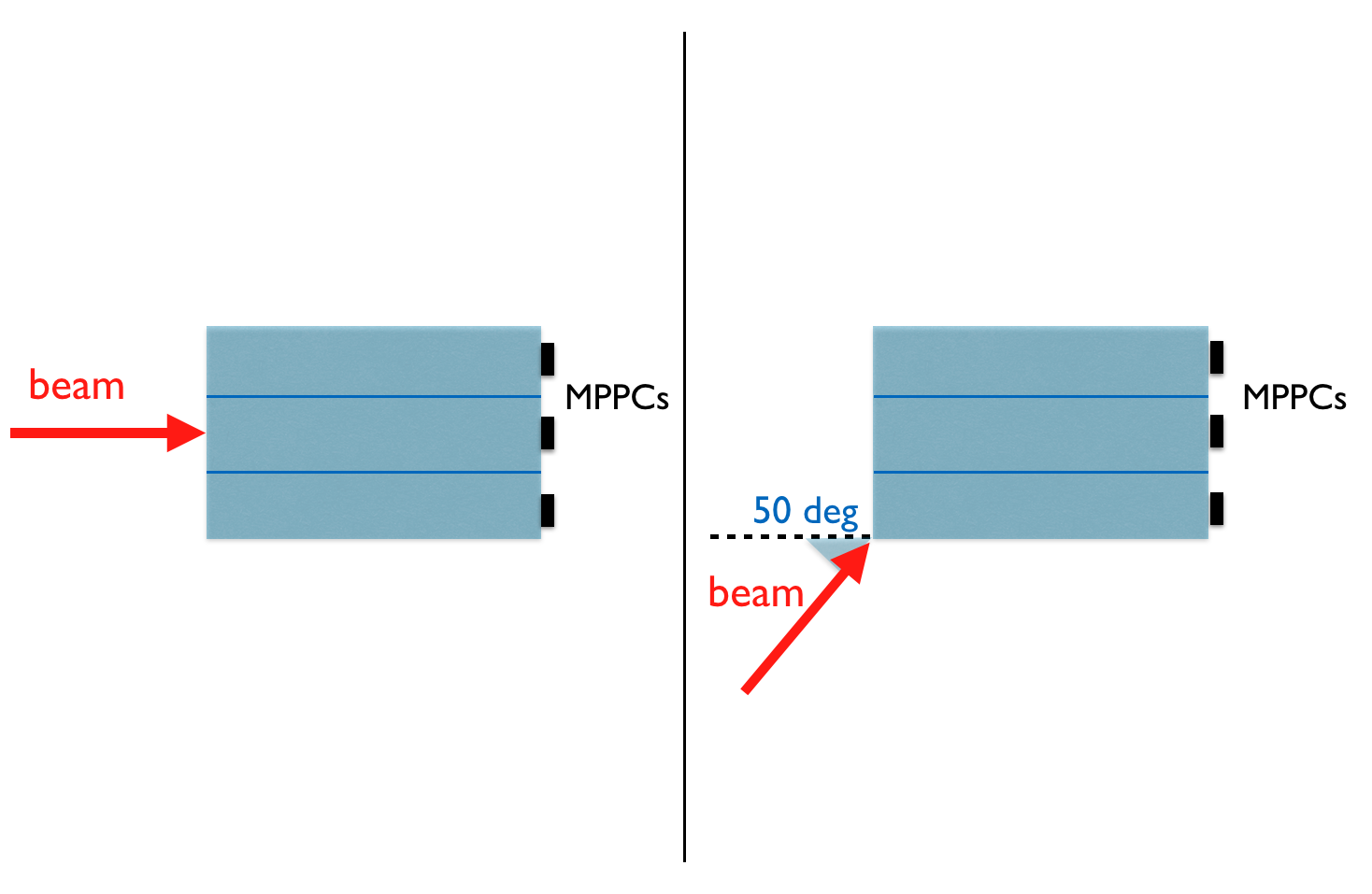}
  \caption{Beam test configurations: beam normal to the prototype
    front face (left) and beam at 50 degrees with respect to the normal of
    the prototype surface (right).}
  \label{fig:configurations}
\end{figure}

\begin{figure}[h!]
\centering
  \includegraphics[width=0.49\textwidth] {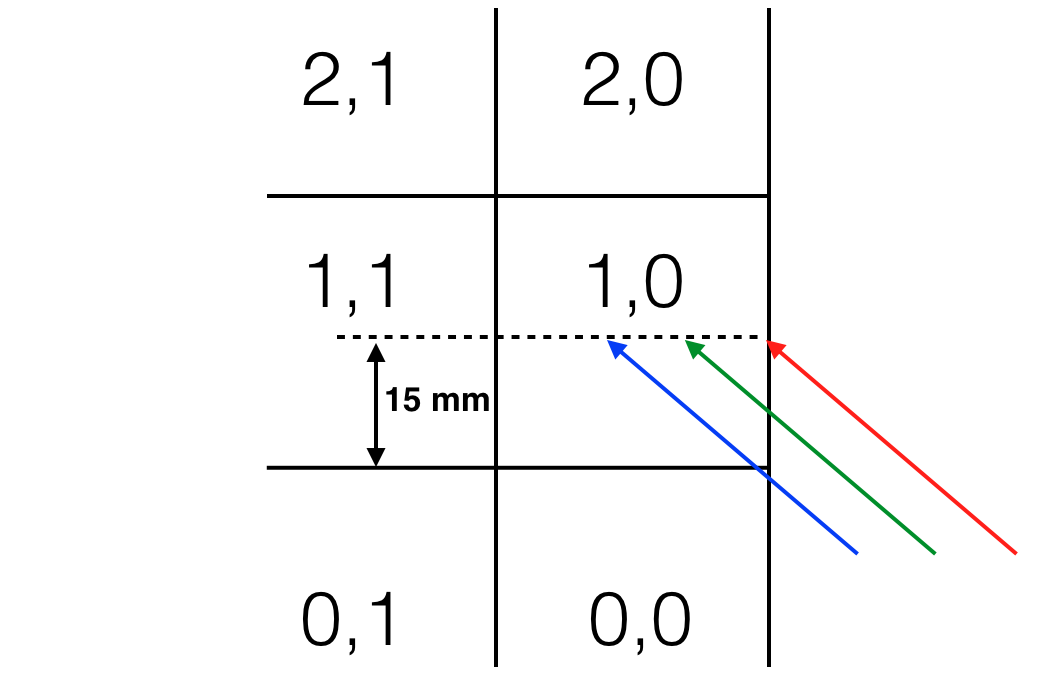}
  \caption{Impinging beam positions used in the run with the beam at
    50 degrees incidence angle. Distance between the impinging points
    is about 7.7 mm. }
  \label{scan_50_deg}
\end{figure}

\begin{figure}[h!]
%%  \centering
  \includegraphics[width=0.49\textwidth] {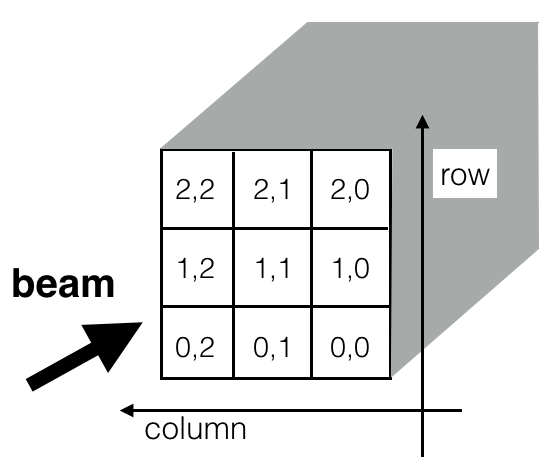}
  \caption{Crystal indexing.}
  \label{fig:crystal-numbering}
\end{figure}

Two configurations, schematically shown in
Figure~\ref{fig:configurations}, were studied during the test:
\begin{enumerate}
\item
  beam at 0 degrees with respect to the prototype front
  face, defined as the side opposite to the photosensors;
\item
  beam at 50 degrees with respect to the prototype surface.
\end{enumerate}
%% The second configuration at 50 \deg represents the most interesting
%% one because this angle is close to the most probable value of the
%% impact angle at the calorimeter for the Mu2e signal
%% electron~\cite{GIANIPEZ}.
Configuration (ii) was motivated by the fact that the expected
average incidence angle of a signal electron  in Mu2e is about 50
degrees.  With the prototype rotated by 50 degrees, data
were taken in three different beam impact points, as shown in
Figure~\ref{scan_50_deg}:
\begin{itemize}
\item 
   on the edge of the crystal (1, 0);
\item
  7.7 mm from the edge;
\item
  15.4 mm from the edge.
\end{itemize}
The channel numbering convention used in the analysis
is shown in Figure~\ref{fig:crystal-numbering}.

The BTF~\cite{Mazzitelli2003524} uses the high current Da$\phi$ne
linac beam to send e$^-$ (e$+$) bunches, with an intensity between
$10^7$ to $10^{10}$ particles/pulse and energy between 300 to 750
(550) MeV, to a Cu target to create secondary low momentum
beams. The target attenuates the intensity and, in cooperation with a
slit system and a bending dipole, allows to select various
configurations of energy and intensity. In our test, we reduced the
intensity to provide an average multiplicity of 0.8 e$^-$ per
pulse in the 80-120 MeV region. The BTF repetition rate is 50 Hz,
with a bunch width of around 10 ns. The beam energy spread is
excellent at high energy but degrades at lower energies. In our
previous test~\cite{LYSOBTF}, we have measured an energy spread of
4.5\% at 100 MeV. For the result reported here the BTF experts 
improved the slit system to reduce this spread as much as possible. We
estimated  the resulting beam energy spread to be less than $2\%$.
\begin{figure}[h!]
%%  \centering
  \includegraphics[width=0.49\textwidth] {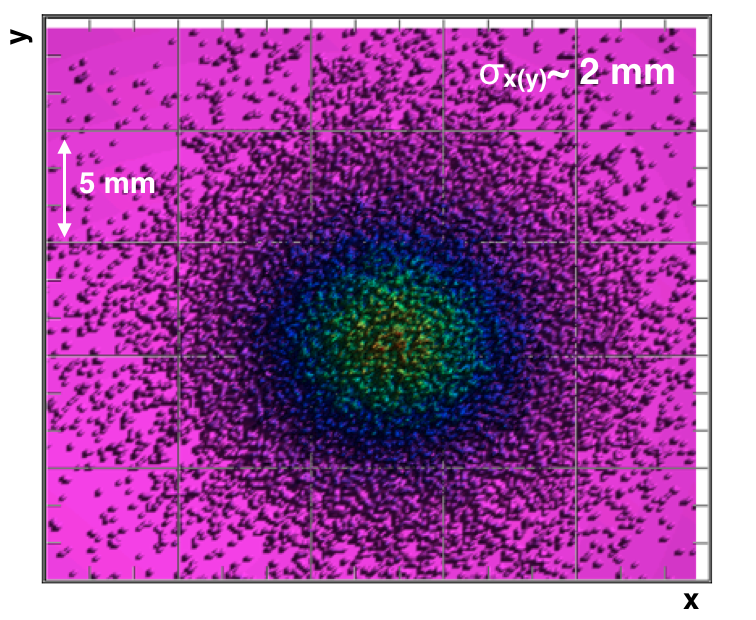}
  \caption{Beam profile.}
  \label{fig:beam-profile}
\end{figure}
The beam divergence in the plane transverse to the beam direction
(X-Y) is small, a Gaussian beam profile is observed with a
$\sigma_{x(y)}$ of about 2 mm as monitored by a dedicated GEM-TPC
system (see Figure~\ref{fig:beam-profile}).

%%%%%%%%%%%%%%%%%%%%%%%%%%%%%%%%%%%%%%%%%%%%%%%%%%%%%%%%%%%%%%%%%%%%%%%%%%%%%%
\section{Charge and time reconstruction}\label{}

A waveform sampling readout with 250 Msps %% allows to design deadtime-less data
%% acquisition system, and is more and more widely used in particle and
%% medical physics. In addition, this kind of system 
allows an accurate analysis of the signals, which is useful in Mu2e to resolve
the pileup of hits.
%% typical problem of high intensity experiments, like the pile-up.
The total charge and the time of the detected pulses were reconstructed %%determined
%%from the digitized waveform 
as follows. 
\subsection{Charge reconstruction}
The charge was estimated by numerical integration of the waveform; two
time windows of 400 ns were used as integration gates in order to
guarantee the full integration of the signals. The first gate was used
to estimate the pedestal $Q_{ped}$ at early time, where no pulses were 
present, while the second gate, around the signal peak, was used to
integrate the signal charge $Q_{signal}$. The reconstructed charge was
then defined as $Q_{reco}= Q_{signal} - Q_{ped}$. The signal time was 
determined by fitting the leading edge of the waveform with an
analytic function.
\subsection{Time reconstruction}
Assuming a constant pulse shape, the best accuracy is 
achieved by setting the signal time at a threshold corresponding to a
constant fraction of the pulse height. Pivotal for this procedure,
usually called digital constant fraction (DCF), are the choices of:
fit function, fit range and threshold. Several parametrizations were
tested: single and double exponential functions, exponential
functions convoluted with a Gaussian, log-normal function
$\exp(-(\ln{x}-\mu)^2/{2 \sigma^2})$, and several others. For each
case, a scan over the DCF threshold and fit range has been
performed.  The fit function that provided the best time resolution was 
the asymmetric log-normal function~\cite{GRUPEN} defined as:
$$
f(t) = N \exp\left( - \frac{\ln^2 \left[ 1 - \eta (t - t_p)/\sigma \right] }{2s_0^2} - \frac{s_0^2}{2} \right) \frac{\eta}{\sqrt{2\pi} \sigma s_0} \ ,
$$
where $N$ is the normalization parameter, $t_p$ is the position of the peak,
$\sigma$ = FWHM/2.35, $\eta$ is the asymmetry parameter, and $s_0$ can
be written as
$$
  s_0 = \frac{2}{\xi} \mbox{arcsinh}  \frac{\eta\xi}{2} , \quad \xi = 2.35\ . 
$$
This can be understood, as the asymmetric log-normal function captures
several important features of the electronic pulse, like the start of the
pulse development at a finite time $t = t_p + \sigma/\eta$,
an exponential growth at a very early stage of the pulse development, and
the presence of the pulse height maximum. %% Waveforms from the photo-sensors of the
%% scintillating counters and of the CsI crystals have different shapes,
%% so the fit range was set differently.

\subsection*{Scintillating counter waveforms}
\begin{figure}[h!]
\centering
  \includegraphics[width=0.49\textwidth]
		  {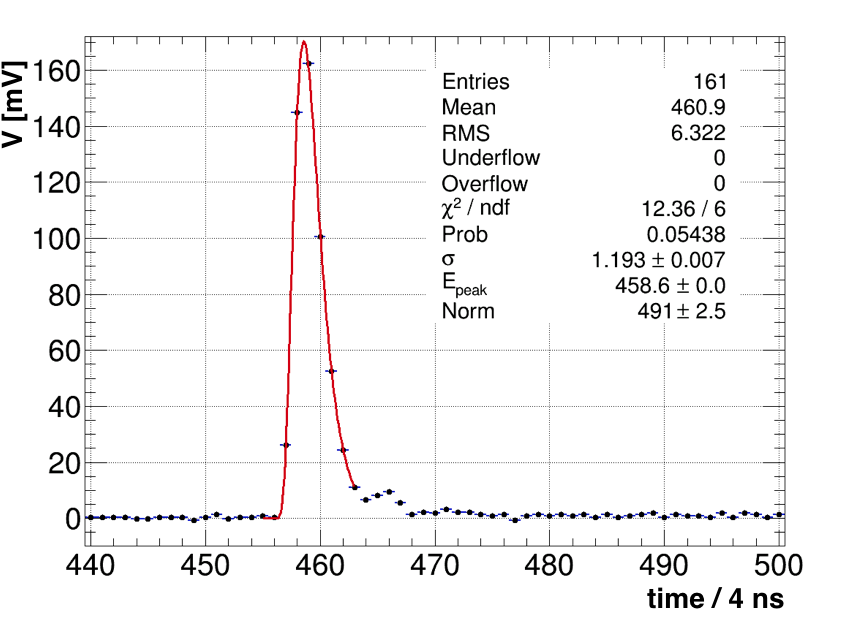}
		  \caption{Fit of a scintillating counter waveform
                    using a log-normal function.}
		  \label{fig:scint-wfs}
\end{figure}

\begin{figure}[h!]
%%\centering
  \hspace*{-0.3cm}
  \includegraphics[width=0.52\textwidth]
		  {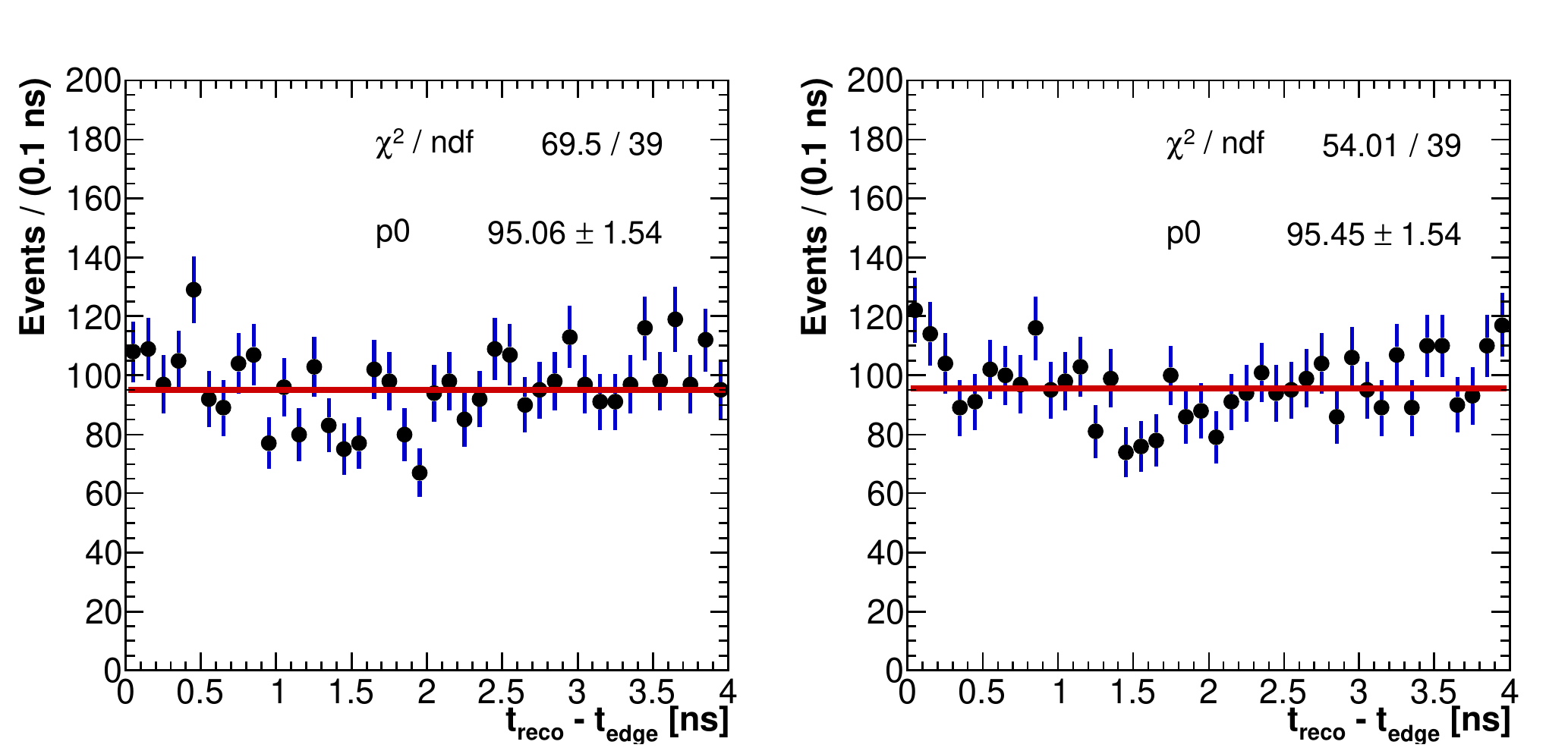}
		  \caption{Distribution of time residuals between the
                    reconstructed time (t$_{\rm reco}$) and the edge
                    of the corresponding digitized sample for the two
                    scintillating counters. Red line shows the fit
                    result to a constant. }

		  \label{fig:scint-wfs-dtbin}
\end{figure}

In the beam test, the scintillating counters had signals with a
leading edge of about 15 ns (3 digitized samples), and a total width
shorter than 100 ns. Since the asymmetric log-normal is defined by 4
parameters, the fit range should be wider than 15 ns. The lower edge
of the fit was set at the first sample where the signal exceeds 5 mV,
while the upper edge was set 16 ns after the peak, thus providing more
than 6 samples to perform the fit. An example of a fit to a pulse is
shown in Figure~\ref{fig:scint-wfs}.
This procedure was checked looking at the distribution of number of
degrees of freedom (\ndof) and $\chi^2/\ndof$.
Presence of systematic effects has been investigated looking at
the distribution of the reconstructed time within the digitized sample
($\Delta \rm t_{\rm edge}$). Figure~\ref{fig:scint-wfs-dtbin} shows
the distributions of $\Delta \rm t_{\rm edge}$ for both scintillating
counters that are reasonably flat.

\subsection*{CsI crystal waveforms}
\begin{figure}[h!]
\centering
  \includegraphics[width=0.49\textwidth]
		  {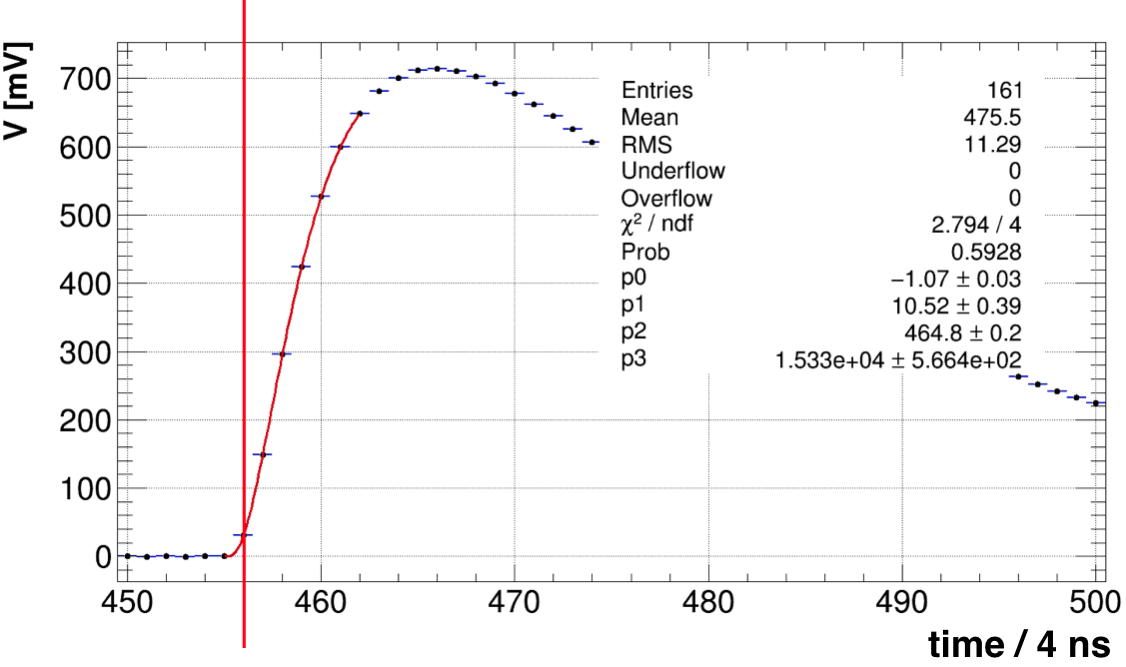}

		  \caption{Example of a fit to a waveform.}
		  \label{fig:crystal-wfs}
\end{figure}

Waveforms corresponding to signals from the CsI crystals convoluted
with the SiPM and pre-amp response function have a leading edge of about
25 ns, and a total width of about 300 ns. The duration of the leading
edge allows to perform the fit only on this signal region. The fit
range has been defined as follows: the lower limit was set at the
first time sample where the pulse exceeds $0.5\%$ of the pulse
maximum, while the upper limit has been set at the first time sample where the
pulse exceeds $85\%$ of the pulse maximum. The DCF threshold, used to
determine the reconstructed time, has been optimized using the data
taken with a 80 MeV electron beam at 0 degrees incidence angle.
\begin{figure}[h!]
\centering
  \includegraphics[width=0.49\textwidth]
		  {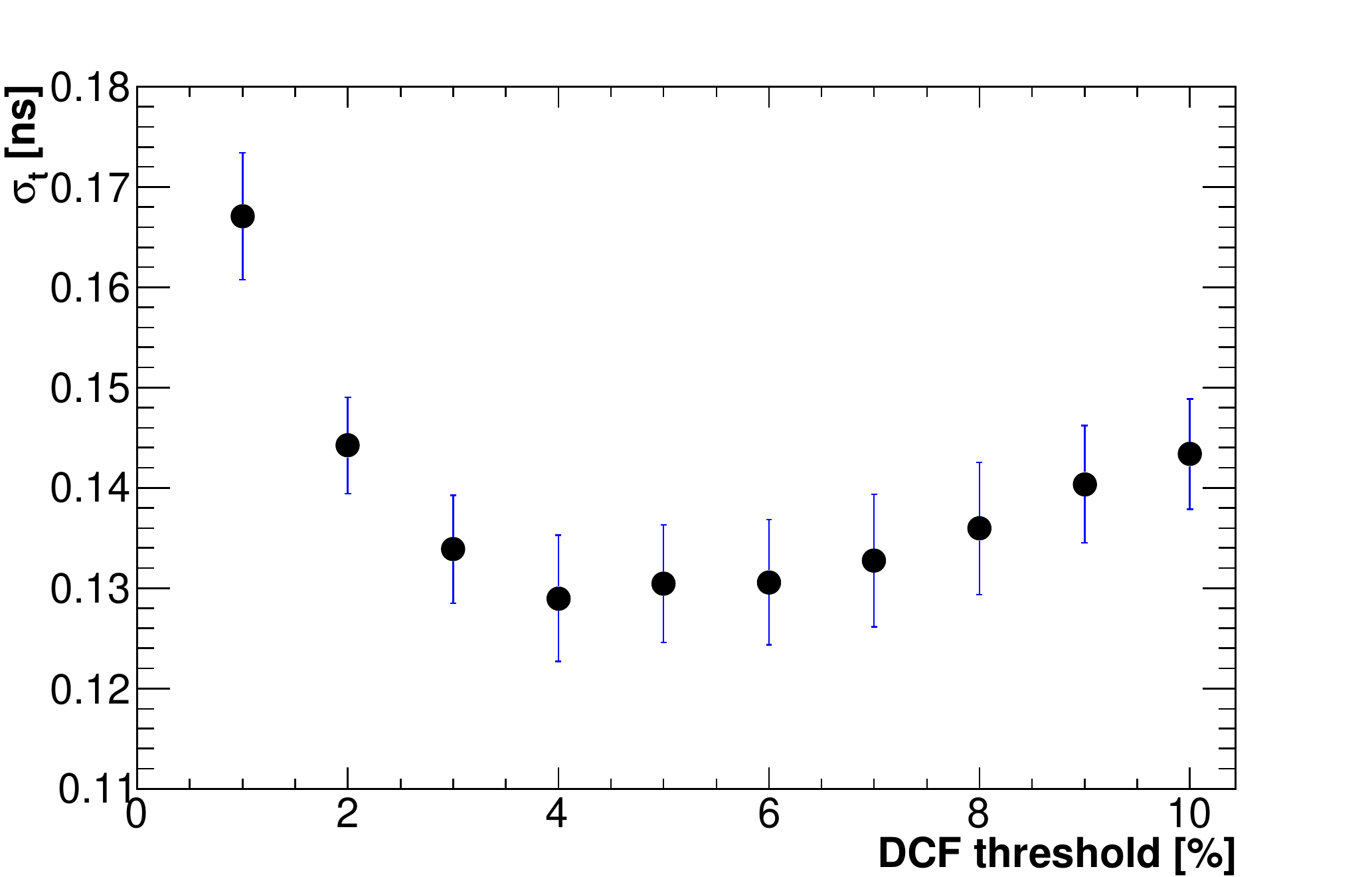}

		  \caption{Time resolution using 80 MeV electron beam
                    as a function of the DCF threshold.}
		  \label{fig:threshold-DCF}
\end{figure}
Figure~\ref{fig:threshold-DCF} shows that, for the thresholds in the
range 2\% - 10\% of the maximal pulse height, the time resolution is
stable within $10\%$. So the DCF threshold has been set to
$5\%$ of the maximum pulse height. Figure~\ref{fig:crystal-wfs} shows
an example of a fit to a CsI crystal waveform.

\begin{figure}[h!]
\centering
    \includegraphics[width=0.49\textwidth]
		  {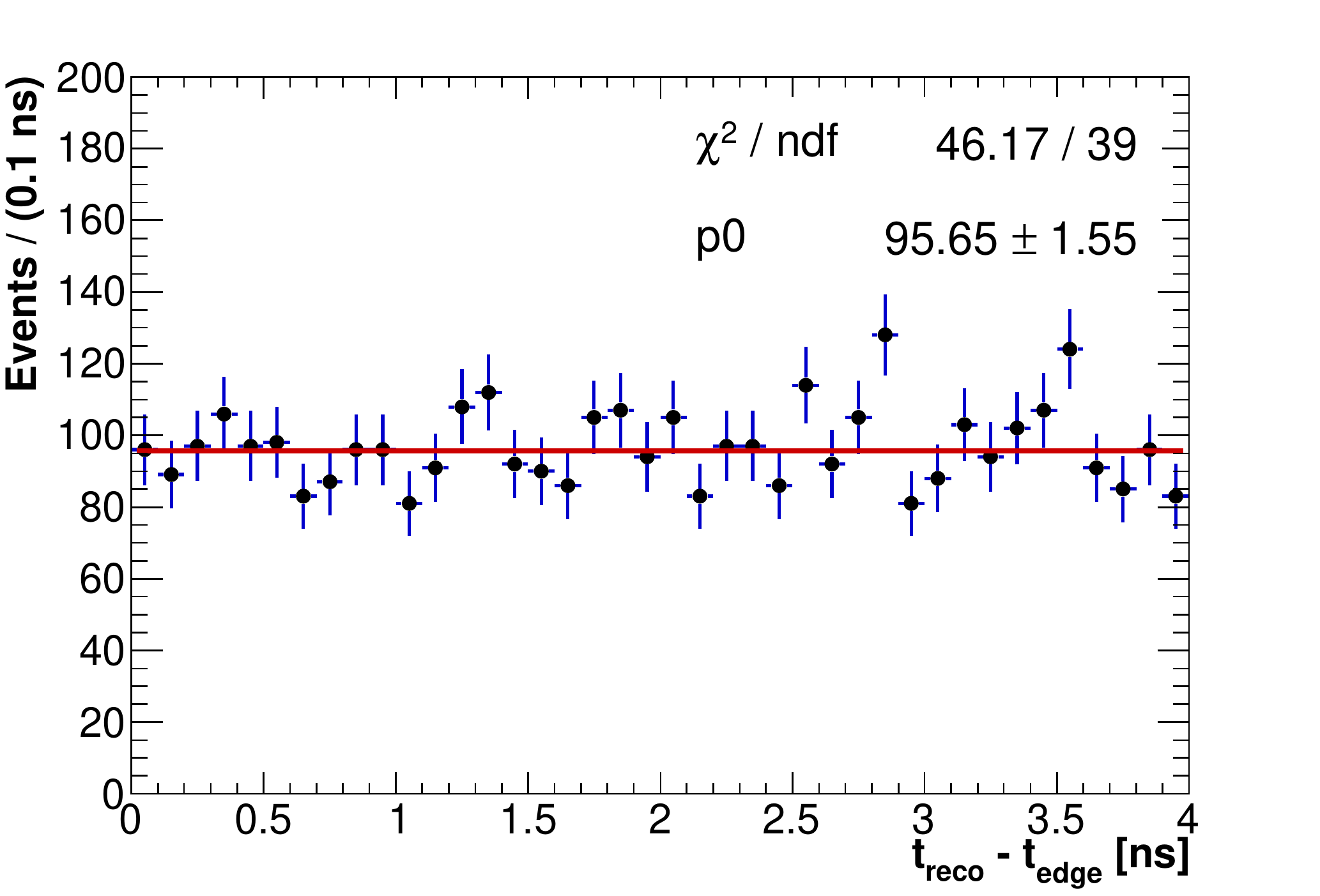}
		  \caption{Distribution of the reconstructed time of
                    the CsI pulses within their respectively digitized
                    sample. Red line shows the fit result to a constant.}
		  \label{fig:crystal-wfs-dtbin}
\end{figure}

The selected range provides more than  7 samples to perform the fit.
The distribution of $\Delta{\rm t_{\rm edge}}$, shown in
Figure~\ref{fig:crystal-wfs-dtbin}, is flat and does not present any
significant systematic effects.

%%%%%%%%%%%%%%%%%%%%%%%%%%%%%%%%%%%%%%%%%%%%%%%%%%%%%%%%%%%%%%%%%%%%%%%%%%%%%%%

\section{Analyses selection}\label{}
Events with a single beam particle within the integration gate were
selected requiring:
%% Beam events for the analysis were selected requiring:
\begin{enumerate}
\item
  Energy deposition in each of the two beam counters consistent with a single
  particle. 
\item
  Pulse-shape discrimination of the waveforms from each of the  CsI crystals
  to discard events with one or more channels saturated because
  of pileup of particles.
\end{enumerate}
\begin{figure}[h!]
  \includegraphics[width=0.49\textwidth]
		  {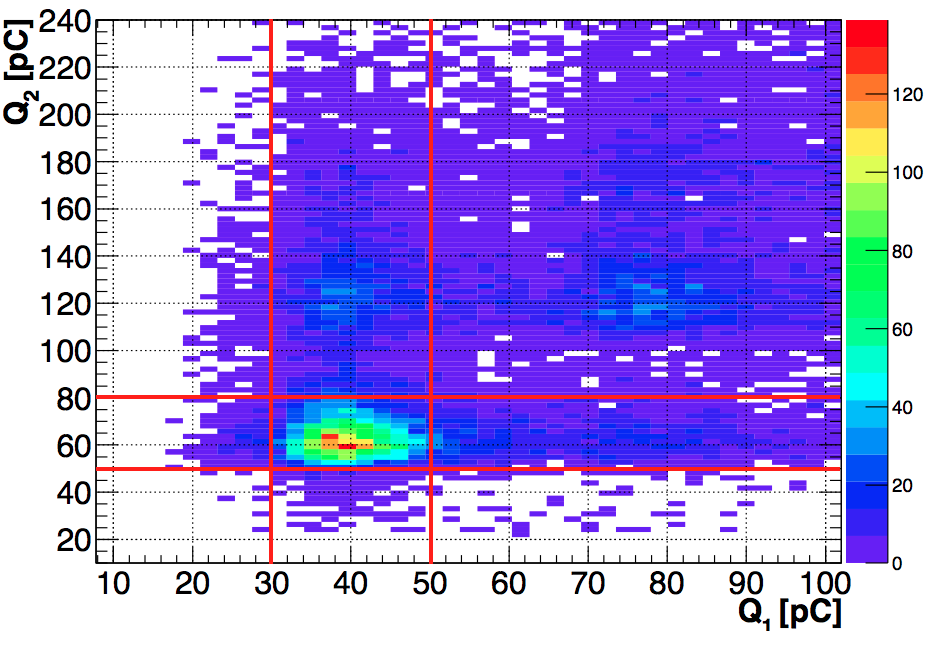}
		  \caption{Scatter plot of the reconstructed charges
                    in the two scintillation counters used for
                    triggering the beam. Red lines represent the
                    applied cuts.}
		  \label{fig:fingers-charges}
\end{figure}

Figure~\ref{fig:fingers-charges} shows the scatter plot of the charges
reconstructed in the beam counters, $\rm Q_1$ and $\rm Q_2$.  The cuts
applied for selecting single particle events are the following:
$\rm Q_1 \in [30, 50]$ pC and $\rm Q_2 \in [50, 80]$ pC. 

\begin{figure}[h!]
  \includegraphics[width=0.49\textwidth]
		  {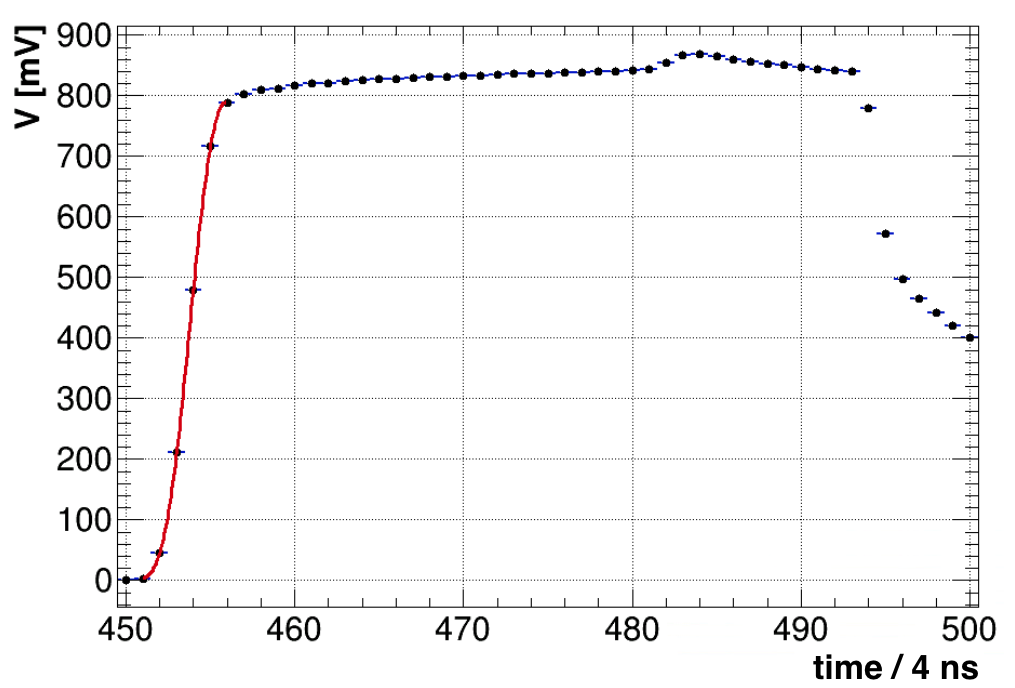}
		  \caption{Example of a saturated signal read out by the channel
                    (0, 0). Red lines show the fit result used for the
                    time reconstruction.}
		  \label{fig:saturated-signals}
\end{figure}

\begin{figure}[h!]
  \includegraphics[width=.49\textwidth]
		  {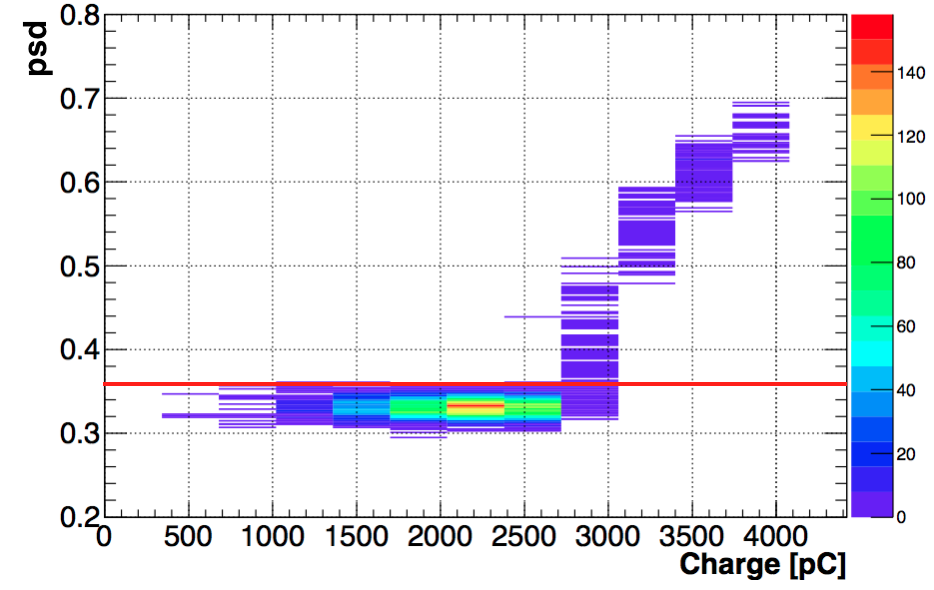}
		  \caption{Pulse shape discriminator variable as a function of 
                    the reconstructed charge. Red line shows the applied cut.}
		  \label{fig:psd}
\end{figure}

However, this selection was not sufficient to discard all the  events with
more than 1 electron, due to the efficiency of the scintillating
counters and their limited
acceptance. Figure~\ref{fig:saturated-signals} shows an example of
a saturated signal not discarded by the cut on $Q_1$ and $Q_2$.
For that reason, an additional selection was made using a
pulse-shape discriminator variable, psd, defined as follows:
$$
  \mbox{psd} = \frac {\int_{a}^{b}\mbox{Waveform}}{\mbox{Total waveform charge} } \ , \nonumber
$$
where $a$ and $b$ correspond to the time samples at $1\%$ of the
maximum pulse height on the leading edge, and $90\%$ of the maximum
pulse height on the trailing edge, respectively. Figure~\ref{fig:psd} shows the 
distribution of psd as a function of the reconstructed
charge for the channel (0, 0). In the following, we consider ``single 
particle events'' as those  with psd $<$ 0.36 on all the reconstructed hits. 
The psd discrimination allows to select a 
clean set of events at each beam energy without directly cutting on the energy 
distribution.
%%  The psd cut has been set at
%% 0.36. Discrimination with the psd variable allows its use at any beam
%% energy without changing the cut value.

%%%%%%%%%%%%%%%%%%%%%%%%%%%%%%%%%%%%%%%%%%%%%%%%%%%%%%%%%%%%%%%%%%%%%%%%%%%%%%%%

\section{Energy and Time calibration}
\begin{figure}[h!]
  \centering 
  \includegraphics[width=0.49\textwidth]
                  {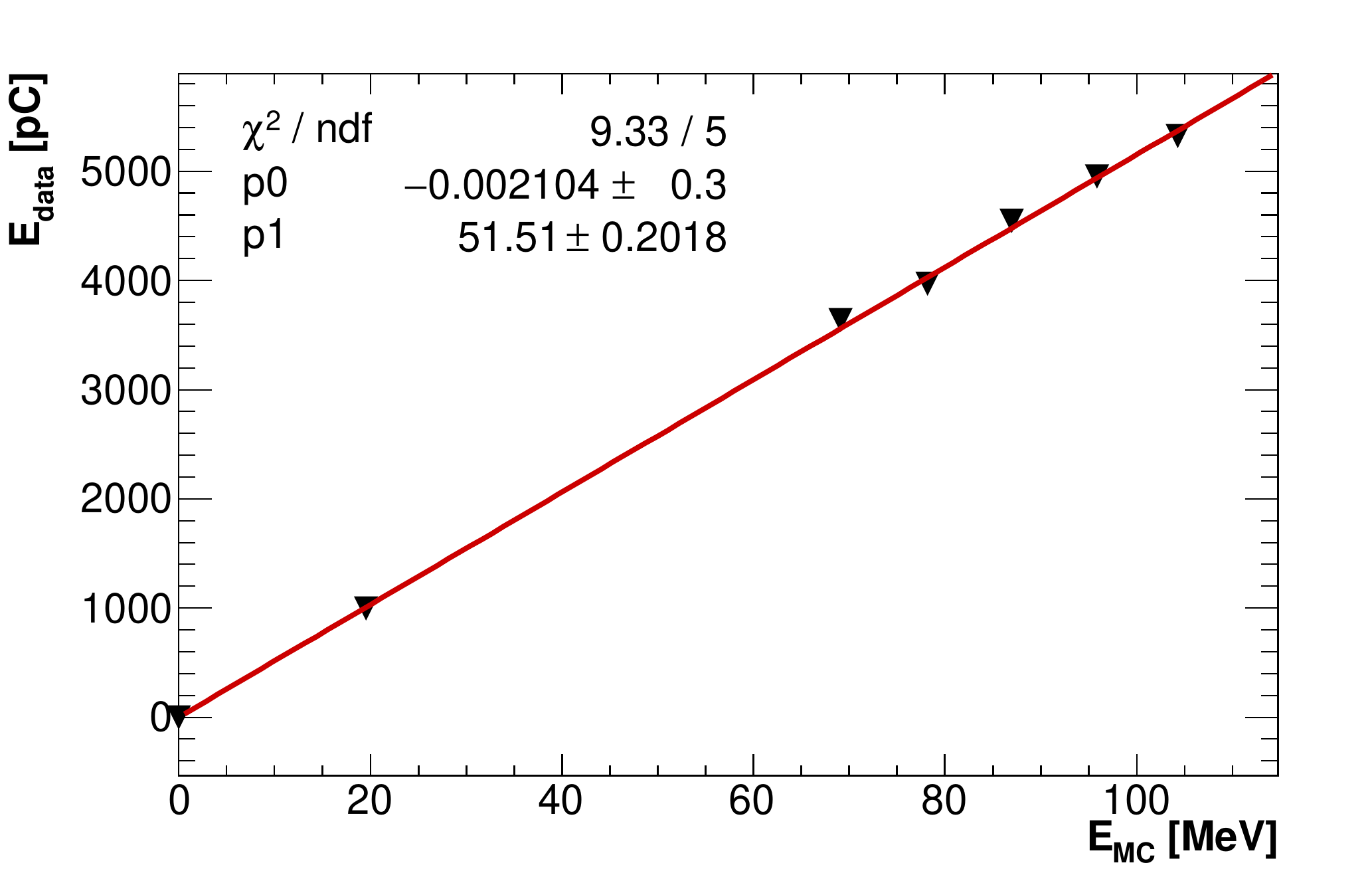}
		  \caption{Reconstructed total charge of the prototype
                    in the calibration runs (E$_{\rm data}$) versus
                    the expected prototype energy deposition from the
                    Monte Carlo simulation (E$_{\rm MC}$).}
		  \label{fig:calibration-curves}
\end{figure}

\begin{figure}[h!]
  \centering
  \includegraphics[width=0.49\textwidth] 
		  {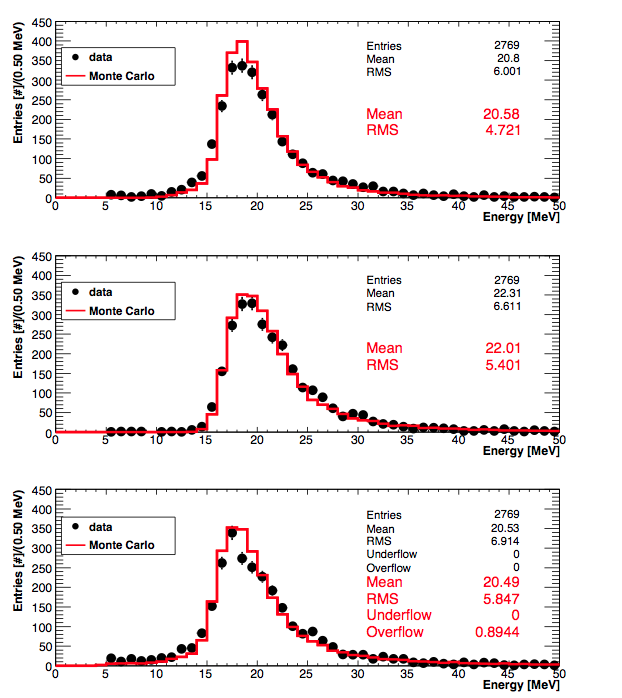}
		  \caption{Energy distributions for the crystals in the central column
                    overlaid with the Monte Carlo for the run with cosmic rays.}
		  \label{fig:cosmic-column-charges}
\end{figure}

Equalization of the time and pulse height responses of the channels was 
achieved using data taken with the 80 MeV $e^-$ beam impinging normally  
on each crystal center.
The energy scale calibration was performed using
the beam at 0 degrees in the energy range [80, 120] MeV, while varying 
the energy in 10 MeV steps. An additional point at the energy of about 20 MeV 
was included using Minimum-Ionizing Particles (MIPs) from the  
cosmic ray data. The energy scale was determined 
comparing data with the results obtained by 
means of a GEANT4~\cite{Agostinelli2003250} 
based Monte Carlo simulation. Figure~\ref{fig:calibration-curves} shows 
the calibration curve with a linear fit superimposed.
The energy response of the calorimeter prototype to the MIPs has also been
compared with the Monte Carlo. Figure~\ref{fig:cosmic-column-charges}
shows the energy distributions of the crystals in the central
column, after calibration and equalization, for data superimposed 
with the Monte Carlo. A good data-Monte Carlo agreement is seen.

\begin{figure}[h!]
  \centering
  \includegraphics[width=0.49\textwidth]
                  {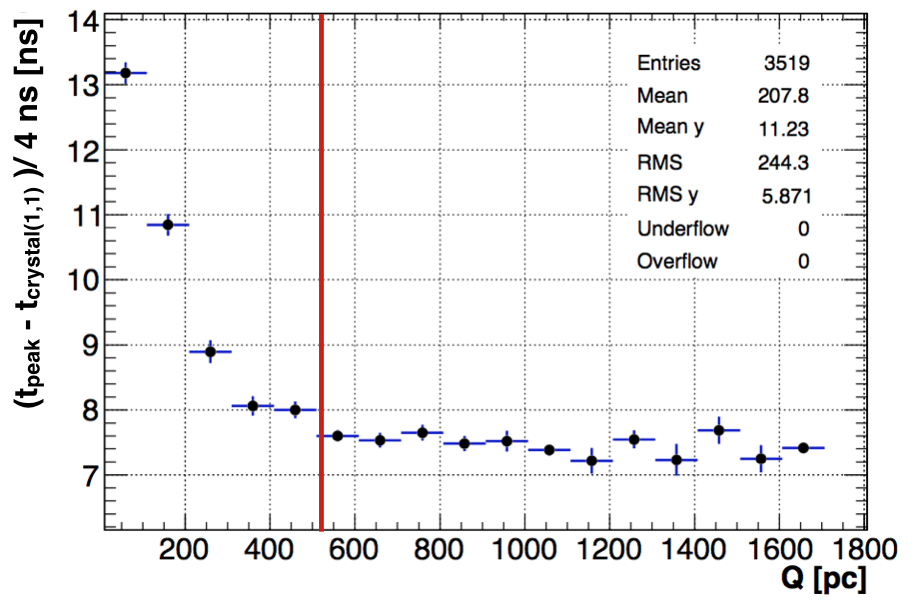} %%dtpeak-example-using-wf-maximum}
	          \caption{Pulse rise time as a function of the reconstructed
                    charge. $\rm t_{\rm peak}$ is the pulse peak time,
                    derived from the log-normal fit. The red line indicates
                    the 10 MeV equivalent threshold.}
	          \label{fig:risingvscharge}
\end{figure}

The same data used for the charge equalization was also  used to determine the
time walk corrections of the prototype channels. A reference time
$\rm t_{scint}$, defined as the average of the beam counter times 
($\rm t_{\rm scint:1,2}$): $\rm t_{\rm scint} = (\rm t_{\rm
  scint:1} + \rm t_{\rm scint:2})/2$, was subtracted from each channel time.
For the beam energies in the range [80, 120] MeV, the jitter of $\rm t_{\rm
  scint}$, defined as the standard deviation of its distribution of
each run, is about ($100 \pm 4$) ps.
The slewing functions of all the prototype channels were applied, 
following the same procedure described in reference~\cite{LYSOBTF}.
Investigating other possible sources of systematic effects, a
dependence of the reconstructed time $\rm t_{\rm crystal}$ on the 
waveform rise time was identified. Figure~\ref{fig:risingvscharge} 
shows the correlation between the rise time and the reconstructed charge.
To reduce the impact of the changing pulse shape on the reconstructed
times, a threshold at 10 MeV was applied to the crystal signals used for
the time resolution studies.

\section{Measurement of the energy resolution}
\begin{figure}[h!]
 \centering
 \includegraphics[width=0.49\textwidth]
 		  {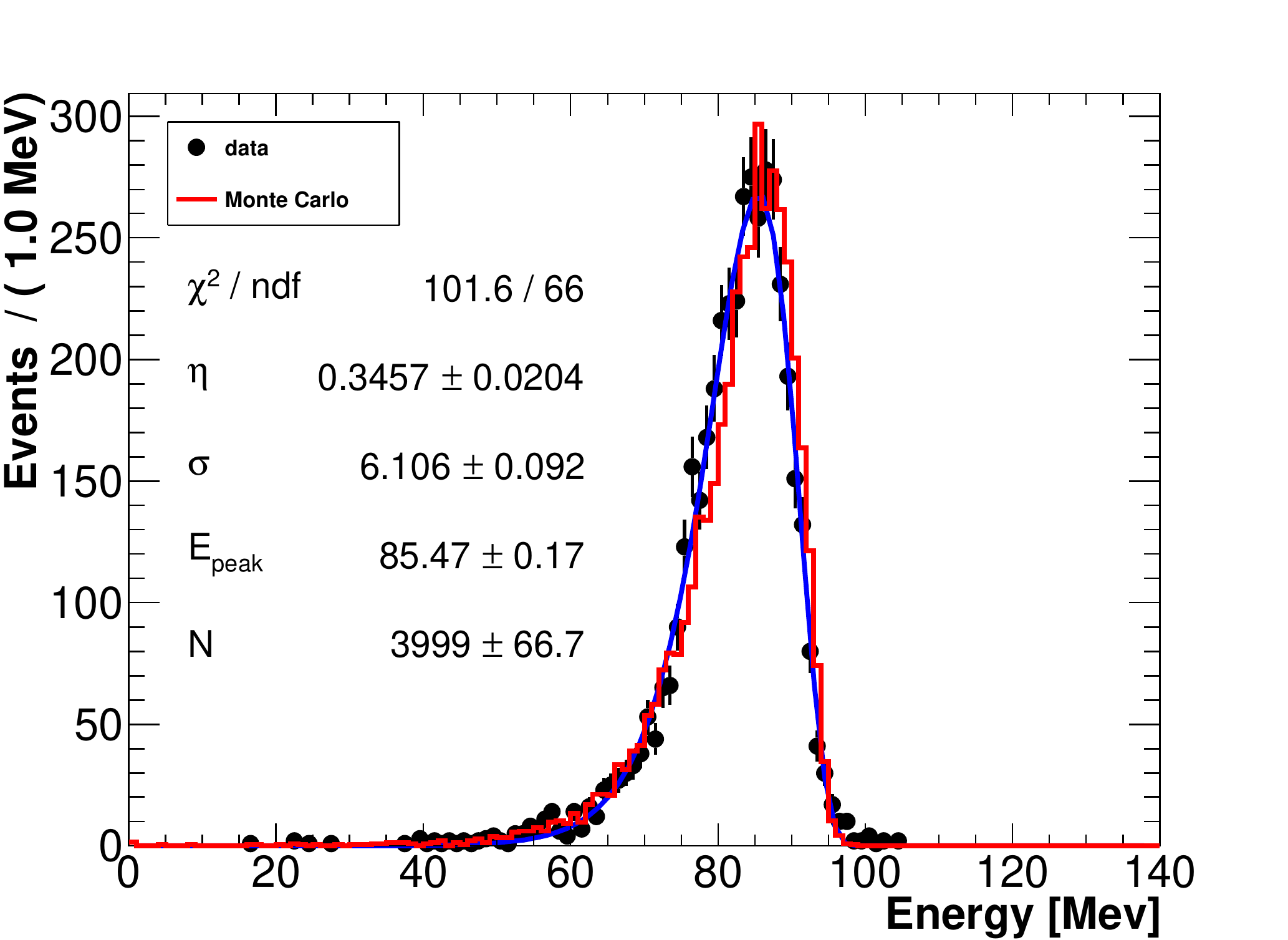}
                  \caption{Distribution of reconstructed energy
                    obtained from the data overlaid with the
                    Monte Carlo for the run with beam energy of 90
                    MeV. Blue line represents a fit to the data with a log-normal function. }
 		  \label{fig:energy-MC-data}
\end{figure}

\begin{figure}[h!]
  \centering
  \includegraphics[width=0.49\textwidth]
		  {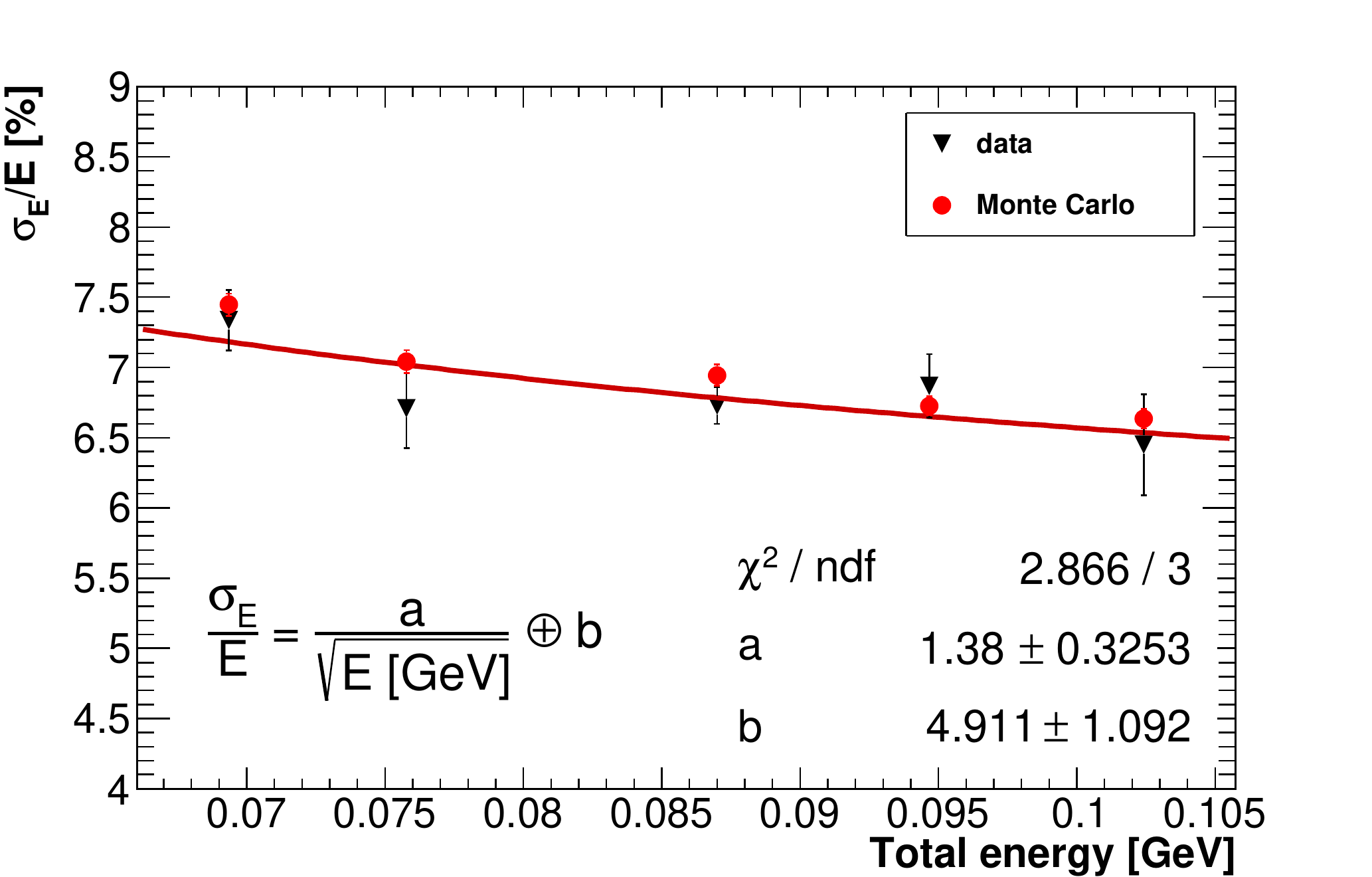} %%energy-resolution-data-vs-mc}
                  \caption{Energy resolution obtained from the data (black) 
                    taken at 0 degrees compared with the Monte Carlo (red).}
		  \label{fig:energy-resol-MC-data}
\end{figure}

\begin{figure}[h!]
 \centering
 \includegraphics[width=0.49\textwidth]
 		  {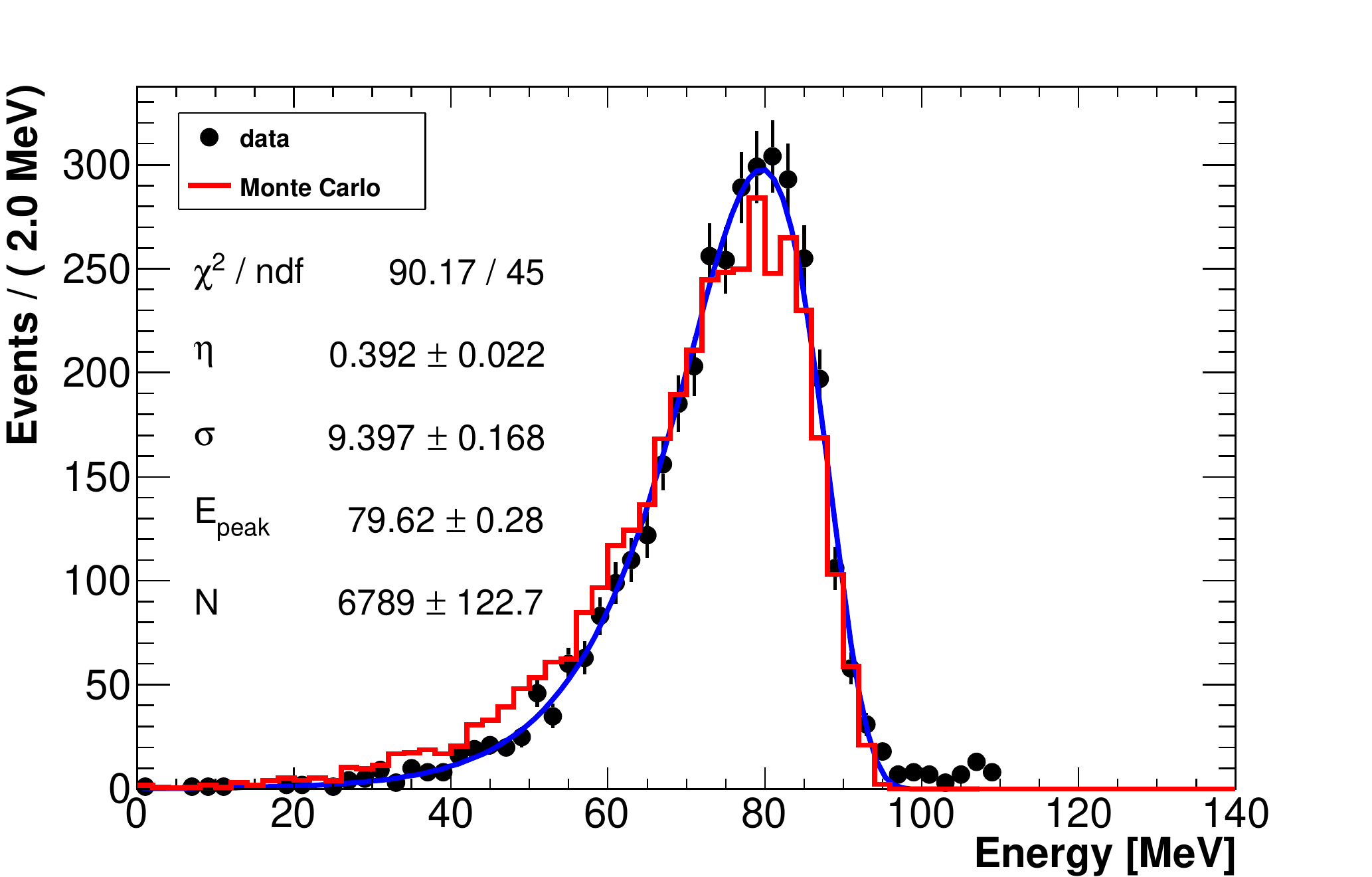}
                  \caption{Distribution of reconstructed energy
                    obtained from the data overlaid with the Monte
                    Carlo for the run with beam energy of 100 MeV and 
                    50 degrees incidence angle. Blue line represents a fit to the data with
                    a log-normal function. }
 		  \label{fig:energy-MC-data-tilted}
\end{figure}

The active volume of the calorimeter prototype was $9\times9\times20$
cm$^3$ and corresponded to $\rm \sim (1.3 \ R_{Moliere})^2 \times( 10 \ X_0)$. 
Due to the small dimensions, the transverse and longitudinal leakages impact
significantly the energy response. 
Figure~\ref{fig:energy-MC-data} shows the distribution of the
total energy deposition obtained from data taken at a beam energy of 
90 MeV and 0 degrees incidence angle compared with the Monte
Carlo. The same Figure shows also a typical fit with a log-normal 
function to the data. The $\sigma$ of this fit was used to evaluate 
the resolution.
Figure~\ref{fig:energy-resol-MC-data} shows the measured energy
resolution as a function of the total energy reconstructed in the
prototype, with the simulation results superimposed. Within the
uncertainties, data and Monte Carlo distributions are in
agreement.%
The measured energy resolution varies from $7.4\%$ to
$6.5\%$ in the energy range [70, 102] MeV.
Figure~\ref{fig:energy-MC-data-tilted} shows that in the configuration
with the beam at 100 MeV and 50 degrees incidence angle the leakage is
larger. Fit results show an energy resolution of about 9 MeV. The same
Figure shows an adequate agreement between data and Monte Carlo.

%%%%%%%%%%%%%%%%%%%%%%%%%%%%%%%%%%%%%%%%%%%%%%%%%%%%%%%%%%%%%%%%%%%%%%%%%%%%%%%

\section{Measurement of the time resolution}
\begin{figure}[h!]
\centering
  \includegraphics[width=0.49\textwidth] {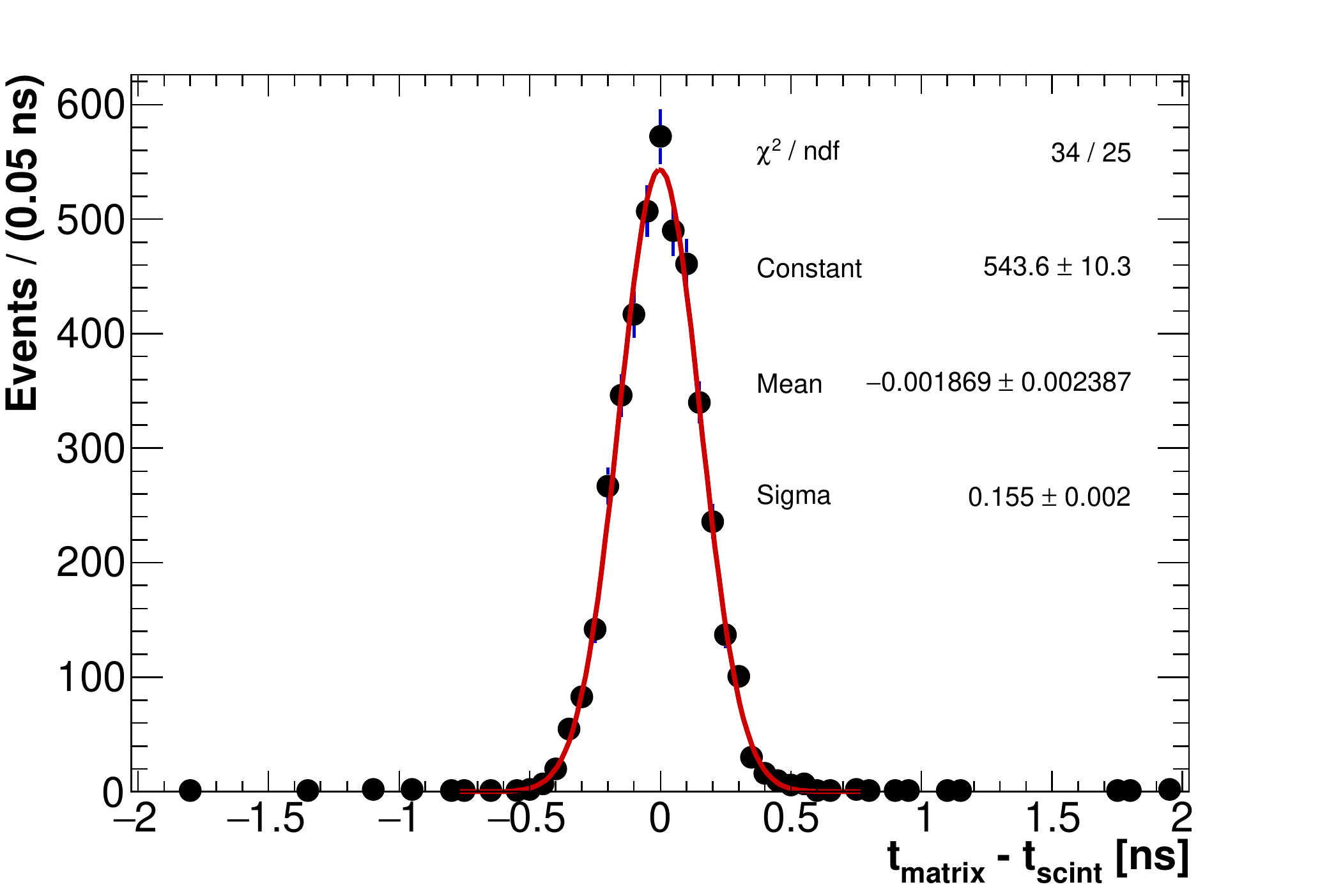}
  \caption{Distribution of time residuals between $\rm t_{matrix}$ and
    $\rm t_{scint}$ for the run at 100 MeV with the beam normal to the
    prototype.}
  \label{dt_matrix_100MeV}
\end{figure}

\begin{figure}[h!]
  \centering
  \includegraphics[width=0.49\textwidth]
		  {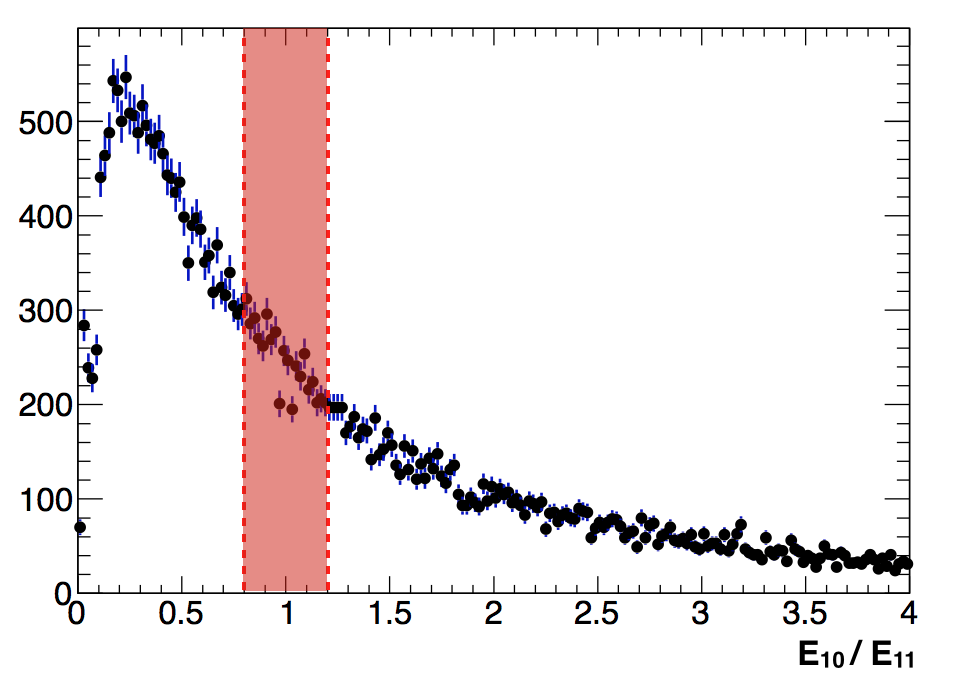}
		  \caption{Distribution of reconstructed energy ratio
                    between crystals (1,0) and (1,1) for the run at
                    100 MeV with the beam impinging at 50 degrees on the
                    prototype.}
		  \label{fig:ratio}
\end{figure}

\begin{figure}[h!]
  \centering
  \includegraphics[width=0.49\textwidth]
		  {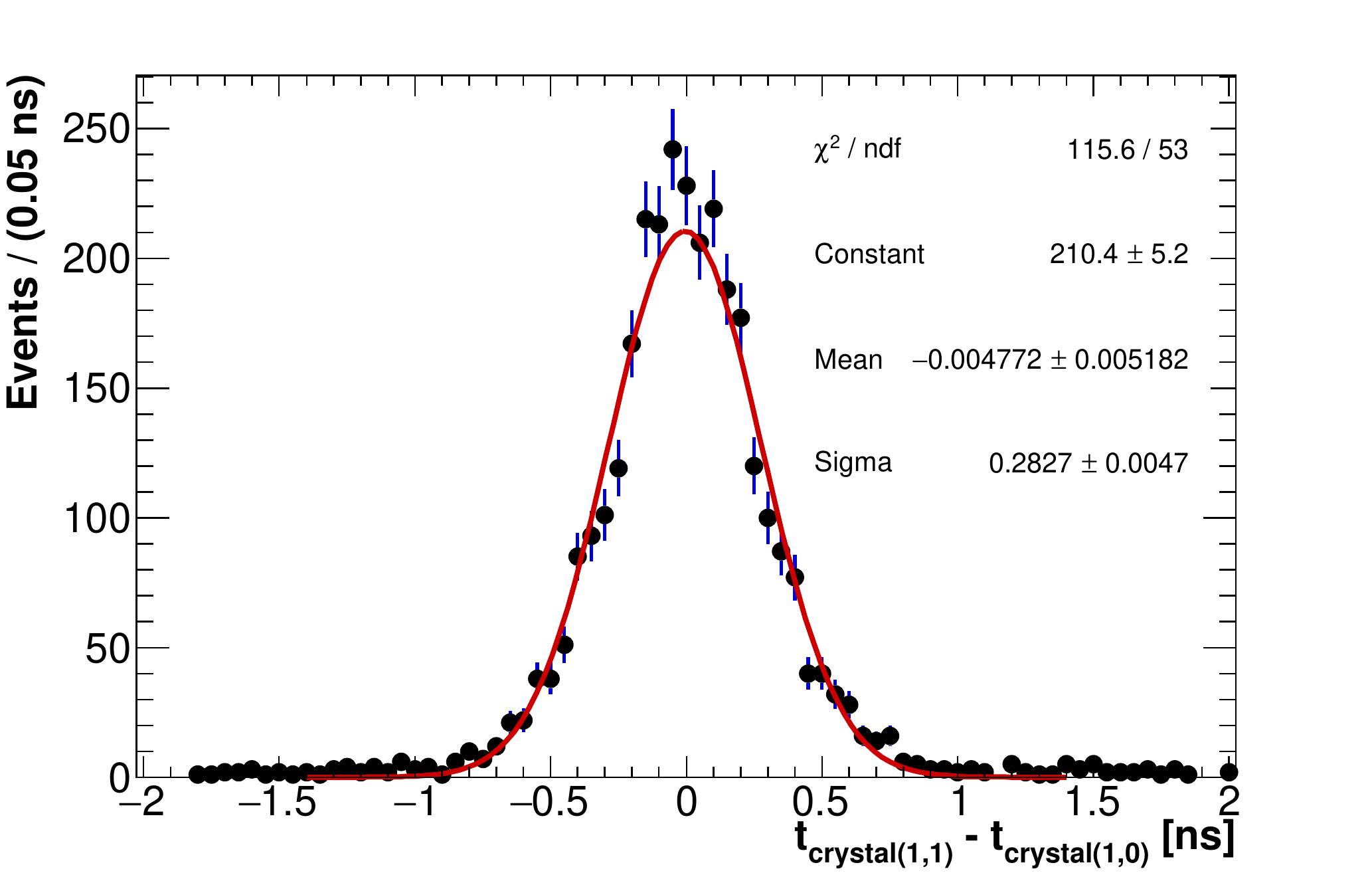}
		  \caption{Distribution of time residuals between
                    channels (1, 0) and (1, 1) for the run at 100 MeV
                    with the beam impinging at 50 degrees on the
                    prototype. }
		  \label{fig:2crystals}
\end{figure}

\begin{figure}[h!]
  \centering
  \includegraphics[width=0.49\textwidth]
		  {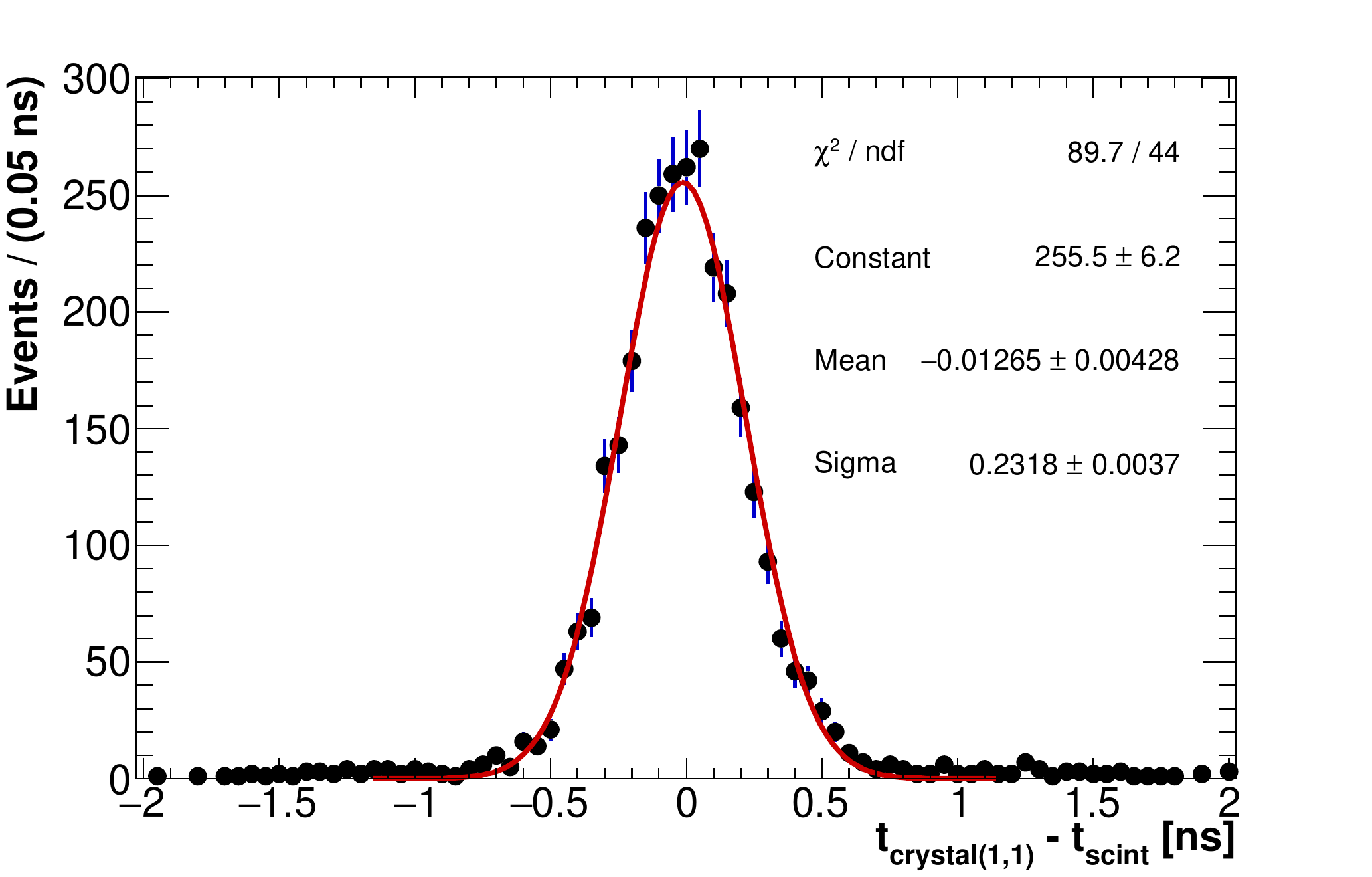}
		  \caption{Distribution of time residuals between $\rm
                    t_{\rm crystal(1, 1)}$ and $\rm t_{\rm scint}$ for
                    the run at 100 MeV with the beam impinging at 50
                    degrees on the prototype.}
		  \label{fig:dtCentralCry50deg}
\end{figure}

The time resolution was measured using three different methods:
\begin{enumerate}
\item[1)]
  using only the crystal with the largest energy deposition;
\item[2)]
  using the energy-weighted mean time of all crystals in the matrix:
$$
  \rm t_{matrix} = \sum_{i,j}{(\rm t_{\rm crystal(i,j)} \cdot \rm E_{i,j})}/E_{\rm tot}, ~~ \rm E_{\rm tot} = \sum_{i,j}{E_{i,j}}\nonumber
$$
\item[3)]
  using two neighboring crystals with similar energy deposition.
\end{enumerate}
The first two techniques require an external time reference $\rm
t_{\rm scint}$. No external time reference is needed for the third one. Methods 1
and 2 were used with both beam configurations: at 0 degrees and 50
degrees. Method 3 was used only for the runs with the beam at 50 degrees, 
because these were the only ones where neighboring crystals with 
reconstructed energies larger than 10 MeV were present.
%
%% \subsection{Beam incidence at 0 $\deg$}
The configuration at 0 degrees represents the simplest one from the
point of view of the analysis, providing a helpful handle for the
development of the time reconstruction method. Figure~\ref{dt_matrix_100MeV} 
shows an example of the distribution of time residual between 
$\rm t_{matrix}$ and $\rm t_{scint}$ for the 100 MeV run.
A Gaussian fit to the same distribution shows a standard deviation 
of about 150 ps, so that removing the contribution of the $\rm t_{scint}$ 
jitter, the resulting time resolution is of about 110 ps.
All data taken in the tilted configurations were combined to apply
Method 3. Crystals (1, 1) and (1, 0) were
used. Figure~\ref{fig:ratio} shows the distribution of the
reconstructed energy ratio $\rm R  = E_{10}/E_{11}$ between the two selected crystals.
To select the events, we required: $0.8 <\rm R < 1.2$. 
Figure~\ref{fig:2crystals} shows the distribution of time
residuals between $\rm t_{\rm crystal(1,1)}$ and $\rm t_{\rm crystal
  (1,0)}$. The sigma resulting from a Gaussian fit to this distribution
is $283$ ps, so that assuming the time resolutions of the two channels to 
be similar, the single channel time resolution is $\sigma_t =
283/\sqrt{2} = 200$ ps.
Varying the cut on R by about 10\% results in no significant 
difference in the time resolution.
%% Figure~\ref{fig:2crystalsvsratio} 
%% shows the dependence of the time resolution with respect to both: 
%% the energy ratio cut, and the mean energy deposited in the crystal 
%% (1, 1) for each selection. Results shown in Figure~\ref{fig:2crystalsvsratio} 
%% includes a systematic error
%% evaluated varying the fit range of the Gaussian fit used to derive the
%% time resolution. The result reflects the physics expectation; as the
%% cut in the energy ratio gets loose, the time resolution deteriorates
%% because of the contribution of the lower energy crystal.
%% \begin{figure}[h!]
%%   \centering
%%   \includegraphics[width=0.49\textwidth]
%% 		  {\figures/dt2Cry-versus-ratio-v2}
%% 		  \caption{Left: time resolution as a function of the
%%                     charge ratio cut. Right: mean energy deposited in
%%                     the central crystal as a function of the cut on $\rm E_{1,1}/E_{1,0}$.}
%% 		  \label{fig:2crystalsvsratio}
%% \end{figure}
%
To cross check the result obtained with this technique, Method 1 was 
used to measure the time resolution in the same
events. Figure~\ref{fig:dtCentralCry50deg} shows the time residual
between $\rm t_{\rm crystal(1, 1)}$ and $\rm t_{\rm
  scint}$. Subtracting in quadrature the $\rm t_{\rm scint}$ jitter of 100 ps
results in a time resolution of about 200 ps that is compatible
with the result obtained with Method 3.
%%%%%%%%%%%%%%%%%%%%%%%%%%%%%%%%%%%%%%%%%%%%%%%%%%%%%%%%%%%%%%%%%%%%%%%%%%%%%%%

\subsection{Cosmic rays}\label{subsec:cosmic-rays}
\begin{figure}[h!]
  \centering
  \includegraphics[width=0.49\textwidth]
		  {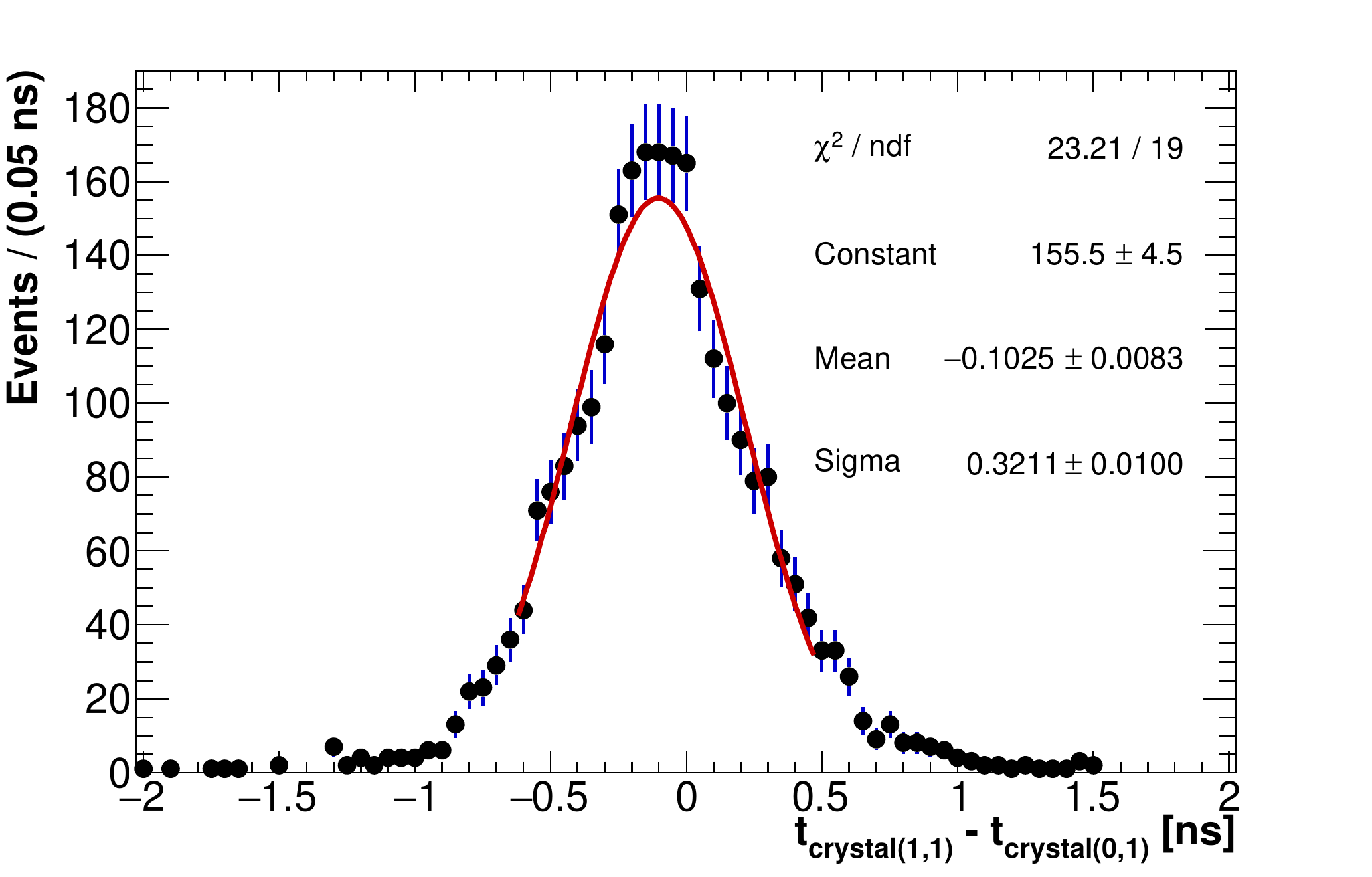}
		  \caption{Distribution of time residuals between
                    crystals (1,1) and (0,1) with Method 3 for the run
                    with cosmic rays. }
		  \label{fig:cosmic-time-resol}
\end{figure}

As MIPs crossing 3 cm of CsI crystal, on
average, deposit about 20 MeV of energy, cosmic muons allow a measurement of
the time resolution in an energy range below the limits of the
BTF. Only events where the cosmic ray crosses the central column
of the prototype were selected, and the ``neighboring crystals''
technique was used to measure the time resolution. This procedure,
however, includes an additional fluctuation due to variations in the
path length of the muons crossing multiple crystals at different
angles.
The cosmic event selection requires a reconstructed energy above 5 MeV
for each of the crystals in the central column, and less than 5 MeV of
deposited energy for each of the other 6 crystals. With a total of 
three crystals in the central column, there are two
independent pairs of neighboring crystals: (2, 1)-(1, 1) and (1,
1)-(0, 1) that were used to measure the time resolution with
Method 3. Figure~\ref{fig:cosmic-time-resol} shows as an example the time
residuals for the pair (1,1) - (1,0). Distributions of time residual were 
fit to a Gaussian function. Then, the time resolution is quoted
using the average of the standard deviations resulting from the two fits. 
Assuming the resolution in all channels to be the same, the resulting 
average is divided by a factor $\sqrt{2}$: $\sigma_{\rm t}\sim 250$ ps.

\begin{figure}[h!]
  \centering
  \includegraphics[width=0.49\textwidth]
		  {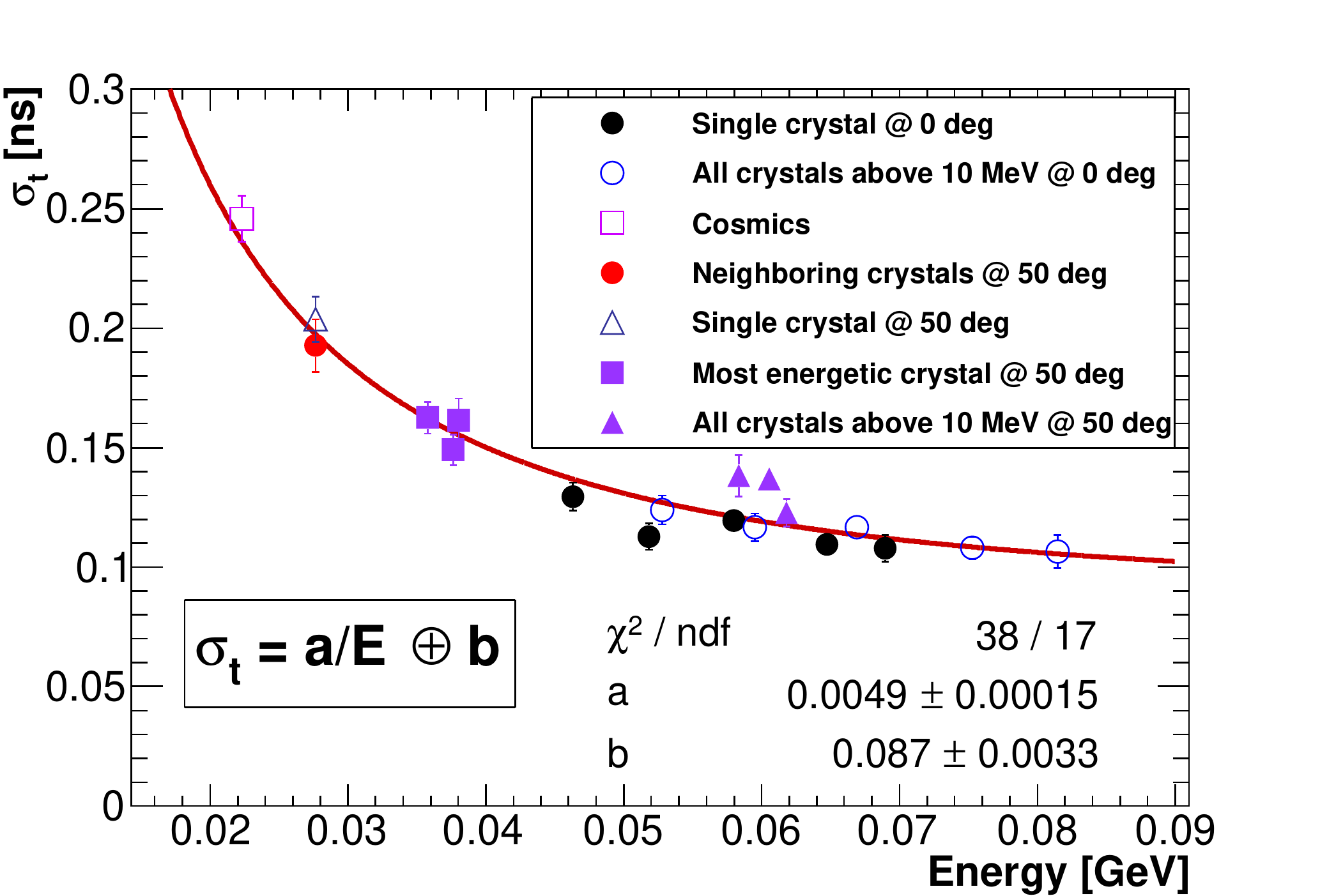}
		  \caption{Time resolution summary plot. }
		  \label{fig:all-single-crystal-and-matrix}
\end{figure}

All results plotted as a function of the energy are summarized in
Figure~\ref{fig:all-single-crystal-and-matrix}. A clear trend of 
the timing resolution dependence on energy is shown. The timing 
resolution ranges from about 250 ps at 22 MeV to about 120 ps in 
the energy range above 50 MeV. The timing resolutions evaluated 
with different methods in the same
energy range are consistent.
Furthermore, in the same energy range, the time resolution using
Method 2 is slightly worse when the beam impacts at 50 degrees (violet
triangles).
Fluctuations of the shower development could result in additional time
jitter between the signals from different crystals and might be
partially responsible for this discrepancy. In principle the time
resolution $\sigma_{\rm t}$ depends on the undoped CsI light emission
characteristics~\cite{CRYO} according to the following formula:
$$
 \sigma_{\rm t} = \frac{a}{E} \oplus b \ , 
$$
where $a$ is proportional to the emission time constant of the undoped 
CsI, and $b$ represents the additional contribute from 
the readout electronics. The fit of the data to this function 
(see Figure~\ref{fig:all-single-crystal-and-matrix}) shows a good agreement 
between the data and the fitted function.

%%%%%%%%%%%%%%%%%%%%%%%%%%%%%%%%%%%%%%%%%%%%%%%%%%%%%%%%%%%%%%%%%%%%%%%%%%%%%%%

\section{Summary}
A reduced scale calorimeter prototype for the Mu2e experiment has been
tested with an electron beam in the energy range [80, 120] MeV at the
Beam Test Facility in Frascati (Italy). Good agreement between data
and Monte Carlo is observed and the measured energy resolution is
dominated by leakage due to the small dimensions of the prototype. The
time resolution $\sigma_t$ as a function of the energy deposition has
been measured using three different techniques that consistently show
that $\sigma_t$ ranges from 250 ps at 22 MeV to about 120 ps above 50
MeV. These results satisfy the Mu2e requirements and also
significantly improve the timing resolution achievable when using
undoped CsI at these low energies.

%%%%%%%%%%%%%%%%%%%%%%%%%%%%%%%%%%%%%%%%%%%%%%%%%%%%%%%%%%%%%%%%%%%%%%%%%%%%%%%%%
%%%%%%%%%%%%%%%%%%%%%%%%%%%%%%%%%%%%%%%%%%%%%%%%%%%%%%%%%%%%%%%%%%%%%%%%%%%%%%%%%
%%%%%%%%%%%%%%%%%%%%%%%%%%%%%%%%%%%%%%%%%%%%%%%%%%%%%%%%%%%%%%%%%%%%%%%%%%%%%%%%%

% use section* for acknowledgement
\section*{Acknowledgment}
The authors express their sincere thanks to the operating staff of the
Beam Test Facility in Frascati (Italy), for providing us a good quality
electron beam, and the technical staff of the participating institutions.
This work was supported by the US Department of Energy; 
the Italian Istituto Nazionale di Fisica Nucleare;
the US National Science Foundation; 
the Ministry of Education and Science of the Russian Federation;
the Thousand Talents Plan of China;
the Helmholtz Association of Germany;
and the EU Horizon 2020 Research and Innovation Program under the 
Marie Sklodowska-Curie Grant Agreement No.690385. 
%

% that's all folks
\end{document}